\documentclass[11pt]{article}
\usepackage{amsmath,amsfonts,amssymb,setspace,comment,url}
\usepackage{geometry}
\usepackage{cite}
\usepackage{xcolor}
\usepackage{graphicx}
\usepackage[small,bf,hang]{caption}
\newcommand{\be}{\begin{equation}}
\newcommand{\ee}{\end{equation}}

\begin{document}

\newcommand{\ben}{\begin{displaymath}}
\newcommand{\een}{\end{displaymath}}
\newcommand{\bea}{\begin{eqnarray}}
\newcommand{\eea}{\end{eqnarray}}
\newcommand{\nn}{\nonumber}
\newcommand{\non}{\nonumber\\}
\newcommand{\bean}{\begin{eqnarray*}}
\newcommand{\eean}{\end{eqnarray*}}
\newcommand{\beqs}{\begin{eqnarray}}
\newcommand{\eeqs}{\end{eqnarray}}

\newcommand{\gM}{{\cal M}}
\newcommand{\gL}{\mathbb{L}}

\newcommand{\SO}[1]{\mathrm{SO}(#1)}
\newcommand{\SU}[1]{\mathrm{SU}(#1)}
\newcommand{\SL}[1]{\mathrm{SL}(#1)}
\newcommand{\ui}[1]{#1}

\newcommand{\mbf}[2]{\mathbf{\left(#1,#2\right)}}

\newcommand{\cT}{{\cal T}}
\newcommand{\trom}{\vartheta}

\newcommand\Tstrut{\rule{0pt}{3ex}}         
\newcommand\Bstrut{\rule[-1.3ex]{0pt}{0pt}}   

\numberwithin{equation}{section}

\begin{titlepage}
	\vfill
	
	\begin{center}
		\baselineskip=16pt
		{\Large \bf Tachyonic Kaluza-Klein modes and the AdS swampland conjecture}
		\vskip 2cm
		{\large \bf Emanuel Malek$^a$\footnote{\tt emanuel.malek@aei.mpg.de}, Hermann Nicolai$^a$\footnote{\tt hermann.nicolai@aei.mpg.de}, Henning Samtleben$^b$\footnote{\tt henning.samtleben@ens-lyon.fr}}
		\vskip .6cm
		{\it $^a$ Max-Planck-Institut f\"{u}r Gravitationsphysik (Albert-Einstein-Institut), \\
			Am M\"{u}hlenberg 1, 14476 Potsdam, Germany \\ \ \\
			\it $^b$ Univ Lyon, Ens de Lyon, Univ Claude Bernard, CNRS,\\
			Laboratoire de Physique, F-69342 Lyon, France \\ \ \\}
		\vskip 1cm
	\end{center}
	\vfill
	
\begin{abstract}
	We compute the Kaluza-Klein spectrum of the non-supersymmetric $\SO{3} \times \SO{3}$-invariant AdS$_4$ vacuum of 11-dimensional supergravity, whose lowest-lying Kaluza-Klein modes belong to a consistent truncation to 4-dimensional ${\cal N}=8$ supergravity and are stable. We show that, nonetheless, the higher Kaluza-Klein modes become tachyonic so that this non-supersymmetric AdS$_4$ vacuum is perturbatively unstable within 11-dimensional supergravity. This represents the first example of unstable higher Kaluza-Klein modes and provides further evidence for the AdS swampland conjecture, which states that there are no stable non-supersymmetric AdS vacua within string theory. We also find 27 infinitesimal moduli amongst the Kaluza-Klein modes, which hints at the existence of a family of non-supersymmetric AdS$_4$ vacua.
\end{abstract}
	\vskip 4cm
	
	\vfill
	
\end{titlepage}

\newpage

\section{Introduction}
The stability of anti-de Sitter (AdS) spacetimes has been a long-standing question in theoretical physics. The question is particularly interesting in the case of non-supersymmetric AdS spacetimes, which are not protected by supersymmetry arguments and corresponding positive mass theorems \cite{Gibbons:1983aq}. In the context of string theory, the fate of non-supersymmetric AdS vacua is especially important. For example, non-supersymmetric AdS vacua provide one of the most explicit ways to apply the AdS/CFT correspondence to QCD or condensed matter systems \cite{Bobev:2011rv}. Moreover, non-supersymmetric AdS compactifications provide a simpler class of non-supersymmetric string solutions than de Sitter vacua, which are time-dependent. Therefore, non-supersymmetric AdS solutions can be seen as a natural stepping stone to understanding de Sitter vacua in string theory.

However, so far no fully-fledged examples of non-supersymmetric but stable AdS vacua in string theory have been constructed. One of the most efficient ways of constructing non-supersymmetric AdS vacua in string theory is by uplifting non-supersymmetric solutions of lower-dimensional gauged supergravities. While many such AdS solutions are known, for example in ${\cal N}=8$ supergravities in four \cite{Warner:1983du,Warner:1983vz,DallAgata:2011aa,Borghese:2012qm,Guarino:2015qaa,Borghese:2013dja,Comsa:2019rcz,Guarino:2019snw} and five dimensions\cite{Khavaev:1998fb,Krishnan:2020sfg,Bobev:2020ttg}, all but a handful of such vacua are already perturbatively unstable within the lower-dimensional supergravity \cite{Gibbons:1983aq,Nicolai:1985hs,Freedman:1999gp,Distler:1999tr,Girardello:1998pd,Bobev:2010ib,Fischbacher:2010ec,Fischbacher:2011jx,Borghese:2013dja,Comsa:2019rcz,Bobev:2020ttg}. That is, typically some of the scalar fields of the lower-dimensional supergravity have masses that violate the Breitenlohner-Freedman (BF) bound \cite{Breitenlohner:1982jf}. Therefore, these scalars are tachyonic and generate an instability. Many non-supersymmetric AdS vacua in string theory are also non-perturbatively unstable \cite{Maldacena:1998uz,Horowitz:2007pr,Narayan:2010em,Ooguri:2017njy,Apruzzi:2019ecr}.

This lack of non-supersymmetric but stable AdS vacua, together with arguments based on a sharpened version of the Weak Gravity Conjecture (WGC) \cite{ArkaniHamed:2006dz}, has recently led to the AdS Swampland Conjecture \cite{Ooguri:2016pdq}, which states that all non-supersymmetric AdS vacua in string theory are unstable. The most prominent possible counterexample is the $\SO{3} \times \SO{3}$-invariant non-supersymmetric AdS$_4$ vacuum of the ${\cal N}=8$ $\SO{8}$ gauged supergravity in four dimensions \cite{Warner:1983du,Warner:1983vz}. This vacuum is perturbatively stable within the full four-dimensional ${\cal N}=8$ supergravity \cite{Fischbacher:2010ec}, which has long prompted hope that it may also be stable when uplifted to the full 11-dimensional supergravity.

Indeed, a perturbative instability would require the masses of the higher Kaluza-Klein modes to drop below those of lowest-lying modes which make up the consistent truncation to ${\cal N}=8$ supergravity and are above the BF bound \cite{Fischbacher:2010ec}. Such behaviour of Kaluza-Klein towers has never previously been observed. Nonetheless, such a \emph{perturbative} instability, therefore, would provide very concrete evidence for the AdS Swampland Conjecture, whose original arguments suggest a non-perturbative instability mechanism. However, since calculating the Kaluza-Klein spectra of supergravity compactifications is a notoriously difficult problem, the question of the perturbative stability of the $\SO{3} \times \SO{3}$-invariant vacuum of 11-dimensional supergravity has remained an open problem thus far.

Recently, a ``brane-jet instability'' was found for the $\SO{3} \times \SO{3}$ vacuum \cite{Bena:2020xxb}. There it is argued that probe branes feel a net repulsive force in certain areas of the compactification manifold due to the varying warp factor of the 11-dimensional solution. This causes the probe branes to be expelled, hence signaling an instability. In \cite{Bena:2020xxb} it is argued that because this instability is localised in the compactification manifold and the Coulomb branch of the probe branes may be reflected in scalars of 4-dimensional supergravity, this brane-jet result might indicate an instability triggered by higher Kaluza-Klein modes.

In this paper, we show that indeed the higher Kaluza-Klein modes of the $\SO{3} \times \SO{3}$-invariant AdS$_4$ vacuum are unstable. We do this by employing the recently developed method \cite{Malek:2019eaz,MalekSamtleben}, which allows us to compute the Kaluza-Klein spectrum of any vacuum of ${\cal N}=8$ gauged supergravity obtained by a consistent truncation from 10-/11-dimensional supergravity, by tracking how the spectrum changes as the vacuum is deformed from the maximally-symmetric one. Here this allows us to follow the spectrum as the round $S^7$, corresponding to the maximally supersymmetric AdS$_4 \times S^7$ vacuum, is deformed to the $\SO{3} \times \SO{3}$-invariant compactification and to compute the Kaluza-Klein spectrum of the $\SO{3} \times \SO{3}$-invariant vacuum up to level 6 above the ${\cal N}=8$ truncation. We find that the higher Kaluza-Klein modes, beginning at level $n = 2$, of the $\SO{3} \times \SO{3}$-invariant AdS$_4$ solution drop below the BF bound and become tachyonic. Therefore, the AdS$_4$ solution is perturbatively unstable in 11 dimensions. Interestingly, this instability is already present at the level of 11-dimensional supergravity and does not require non-perturbative effects of M-theory.

\section{Kaluza-Klein spectroscopy} \label{s:Spectroscopy}
We begin by reviewing the result of \cite{Malek:2019eaz,MalekSamtleben}, which we will use to compute the Kaluza-Klein spectrum of the $\SO{3} \times \SO{3}$-invariant AdS$_4$ vacuum. In \cite{Malek:2019eaz,MalekSamtleben}, Exceptional Field Theory (ExFT) \cite{Hohm:2013pua} was used to derive mass formulae for the Kaluza-Klein spectrum of any vacuum obtained by a consistent truncation to a lower-dimensional maximally supersymmetric supergravity.

ExFT provides a convenient reformulation of 10-/11-dimensional supergravity, which unifies gravitational and flux degrees of freedom in a way that makes an $E_{7(7)}$ symmetry manifest. In particular, the bosonic sector consists of a four-dimensional metric $g_{\mu\nu}$, a generalised metric $\gM_{MN}$ parameterising the coset space $E_{7(7)}/\SU{8}$, and a four-dimensional gauge field $A_\mu{}^M$ transforming in the $\mathbf{56}$ of $E_{7(7)}$, with $\mu = 0, \ldots, 3$ and $M = 1, \ldots, 56$. As shown in \cite{Malek:2019eaz,MalekSamtleben}, in ExFT the Kaluza-Klein fluctuations of any vacuum of a given ${\cal N}=8$ gauged supergravity, whose uplift to 10-/11-dimensional supergravity is known, can be conveniently expressed as a product of the lowest-lying modes, which make up the consistent truncation to the ${\cal N}=8$ supergravity, with the scalar harmonics of the compactification manifold, ${\cal Y}^{\Sigma}$, \emph{at the maximally symmetric point}. For example, for all vacua within the four-dimensional $\SO{8}$ gauged supergravity, ${\cal Y}^{\Sigma}$ would be the scalar harmonics of the round $S^7$.

In this manner, the scalar fluctuations are parameterised by a tensor product of an $\mathfrak{e}_{7(7)} \ominus \mathfrak{su}(8)$-valued matrix, corresponding to the ${\cal N}=8$ gauged supergravity modes, with the scalar harmonics, which we label as above by $\Sigma$. Thus we represent the scalar fluctuations by $j_{A}{}^{B}{}_{\Sigma}$, where $A, B = 1, \ldots, 56$, and which, for fixed $\Sigma$, is valued in $\mathfrak{e}_{7(7)} \ominus \mathfrak{su}(8)$, i.e.
\begin{equation}
 \begin{split} \label{eq:ScalarFluct}
  \mathbb{P}_{A}{}^{B}{}_{C}{}^{D} j_{C}{}^{D}{}_{\Sigma} &= j_A{}^{B}{}_{\Sigma} \,, \\
  j_{A}{}^{C}{}_{\Sigma}\, \delta_{BC} &= j_{B}{}^{C}{}_{\Sigma}\, \delta_{AC} \,,
 \end{split}
\end{equation}
where $\mathbb{P}_{A}{}^{B}{}_{C}{}^{D}$ is the projector onto the adjoint of $E_{7(7)}$. From here onwards, we will freely raise/lower all $A, B$ and $\Sigma, \Omega$ indices by $\delta_{AB}$ and $\delta_{\Sigma\Omega}$, respectively. Since the four-dimensional scalar fields of any 10-/11-dimensional background parameterise the coset space $E_{7(7)}/\SU{8}$, a general linear fluctuation is parameterised by an element of the Lie algebra $\mathfrak{e}_{7(7)} \ominus \mathfrak{su}(8)$. Moreover, we can expand these fluctuations in terms of any complete basis of functions, such as the scalar harmonics of the maximally symmetric compactification, ${\cal Y}^\Sigma$. This leads to the completely general parameterisation of any linear scalar fluctuations of a background in ExFT in terms of the $j_{AB,\Sigma}$ as
\begin{equation}
 \begin{split} \label{eq:ScalarFluctAnsatz}
  \gM_{MN} &= U_{M}{}^{A}\,U_{N}{}^{B}\, \left( \delta_{AB} + \sum_\Sigma \,{\cal Y}^\Sigma\,j_{AB,\Sigma} \right) \,,
 \end{split}
\end{equation}
where the sum runs over the scalar harmonics and $U_M{}^A \in E_{7(7)}/\SU{8}$ is the generalised vielbein describing the 10-/11-dimensional background.

The method of \cite{Malek:2019eaz,MalekSamtleben} gives the mass matrices of the Kaluza-Klein modes in terms of four-dimensional ${\cal N}=8$ supergravity data and information about the uplift to 10-/11-dimensions \emph{at the maximally symmetric point} of the four-dimensional supergravity. In particular, the effect of the deformation away from the maximally symmetric solution on the Kaluza-Klein masses can be deduced entirely from the four-dimensional information. This four-dimensional information consists of
\begin{itemize}
	\item the embedding tensor of the ${\cal N}=8$ gauged supergravity, $X_{\ui{MN}}{}^{\ui{P}}$ with $\ui{M}, \ui{N} = 1, \ldots, 56$,
	\item the four-dimensional scalar matrix ${\cal V}_{A}{}^{\ui{M}} \in E_{7(7)}/\SU{8}$ of the vacuum we want to study.
\end{itemize}
The conventions for the four-dimensional scalar manifold are such that ${\cal V}_A{}^{\ui{M}} = \delta_A{}^{\ui{M}}$ corresponds to the maximally symmetric point, e.g. the round $S^7$ for the $\SO{8}$ gauged supergravity.

We also need the following higher-dimensional data:
\begin{itemize}
	\item the scalar harmonics, ${\cal Y}^\Sigma$, of the compactification,
	\item the linear action of the Killing vector fields of the compactification on the scalar harmonics, which we denote by $\cT_{\ui{M}}{}^{\Sigma}{}_{\Omega}$.\footnote{For non-compact gaugings, we need to know the action of a different set of vector fields, which are not all Killing vectors. The full details are given in \cite{Malek:2019eaz,MalekSamtleben} but are not important here since we are working with the $\SO{8}$ gauged supergravity.}
\end{itemize}
To be explicit, $\cT_{\ui{M}}{}^{\Sigma}{}_{\Omega}$ is defined as
\begin{equation} \label{eq:TMatrix}
 L_{{\cal K}_{\ui{M}}} {\cal Y}^\Sigma = - \cT_{\ui{M}}{}^{\Sigma}{}_{\Omega}\, {\cal Y}^\Omega  \,,
\end{equation}
where ${\cal K}_{\ui{M}}{}$ are the Killing vectors of the compactification\footnote{With appropriate modification for non-compact gaugings as discussed in \cite{Malek:2019eaz,MalekSamtleben}.} and $L$ denotes the Lie derivative. Therefore, the matrices $\cT_{\ui{M}}{}^{\Sigma}{}_{\Omega}$ correspond to the generators of the gauge group (generated by the Killing vectors ${\cal K}_M$) in the representation of the scalar harmonics ${\cal Y}_\Sigma$. These are normalised relative to the four-dimensional embedding tensor, $X_{MN}{}^P$, such that
\begin{equation} \label{eq:TNorm}
 \left[ \cT_M,\, \cT_N \right] = X_{[MN]}{}^{P}\, \cT_P \,.
\end{equation}
Just as in 4-dimensional gauged supergravity in the embedding tensor formalism, the indices $M, N, P = 1, \ldots, 56$ here count electric and magnetic vector fields in order to make $E_{7(7)}$ manifest. However, at most 28 of the electric/magnetic vector fields can contribute to the gauging of the supergravity, and thus at most 28 of $\cT_M$'s can be non-zero. For example, as elaborated on in section \ref{s:Harmonics}, for the $\SO{8}$ gauged supergravity, we can decompose the $\cT_M$ with respect to $\SL{8} \subset E_{7(7)}$, such that
\begin{equation}
 \mathbf{56} \longrightarrow \mathbf{28} \oplus \mathbf{28}' \,.
\end{equation}
Then
\begin{equation}
 \cT_M{}^{\Sigma}{}_\Omega = \left( \cT_{ab},\, \cT^{ab} \right)^\Sigma{}_\Omega \,,
\end{equation}
with $\cT_{ab}{}^\Sigma{}_{\Omega}$ corresponding to the $\SO{8}$ generators and $\cT^{ab\,\Sigma}{}_{\Omega} = 0$.

We emphasise once more that we only require the higher-dimensional information for the maximally-symmetric compactification, e.g. the round $S^7$ for the $\SO{8}$ gauged supergravity, irrespective of the four-dimensional vacuum that we are studying. The power of the method developed in \cite{Malek:2019eaz,MalekSamtleben} is that the effect of the deformation away from the maximally symmetric point can be captured simply by dressing the four-dimensional embedding tensor $X_{\ui{MN}}{}^{\ui{P}}$ and the generators $\cT_{\ui{M}}{}^{\Sigma}{}_{\Omega}$ by the four-dimensional scalar matrix ${\cal V}_A{}^{\ui{M}}$.

In particular, the scalar mass matrix, $M^{AB\Sigma,CD\Omega}$ for the scalar fluctuations, $j_{AB,\Sigma}$, is given by
\begin{equation}
\begin{split} \label{eq:ScalarMassMatrix}
M^{AB\Sigma,CD\Omega}\, j_{AB,\Sigma}\, j_{CD,\Omega} &= \frac17 \left[ X^{ACD} X^{B}{}_{CD} + 2\, X^{CAD} X_{C}{}^{B}{}_{D} + 7 X^{A}{}_{C}{}^D X^{B}_D{}^C \right] j_{A}{}^{E,\Sigma} j_{BE,\Sigma} \\
& \quad + \frac27 \left[ 2\, X^{ACE} X^{BD}{}_{E}  - X^{EAC} X_{E}{}^{BD} \right] j_{AB}{}^{\Sigma}\, j_{CD,\Sigma} \\
& \quad - 4 \, X^{ACD}\, \cT^{B,\Omega \Sigma} \, j_{AB,\Sigma}\, j_{CD,\Omega} - 4 \, X^{CAB}\, \cT_{C}{}^{\Omega}{}_{\Sigma}\, j_{A}{}^{D,\Sigma}\, j_{BD,\Omega} \\
& \quad + 24\, \cT_{A}{}^{\Omega}{}_{\Lambda}\, \cT^{B,\Lambda}{}_{\Sigma}\, j^{AC,\Sigma} \, j_{BC,\Omega} - \cT^{C,\Omega}{}_{\Lambda}\, \cT_{C}{}^{\Lambda}{}_{\Sigma}\, j^{AB,\Sigma}\, j_{AB,\Omega} \,,
\end{split}
\end{equation}
where
\begin{equation}
 \begin{split}
  X_{AB}{}^C &= {\cal V}_{A}{}^{\ui{M}}\, {\cal V}_B{}^{\ui{N}}\, ({\cal V}^{-1})_{\ui{P}}{}^C\, X_{\ui{MN}}{}^{\ui{P}} \,, \\
  \cT_A{}^{\Sigma}{}_{\Omega} &= {\cal V}_{A}{}^{\ui{M}}\,\cT_{\ui{M}}{}^{\Sigma}{}_{\Omega} \,,
 \end{split}
\end{equation}
are the embedding tensor and linear action of Killing vectors dressed by the scalar matrix, ${\cal V}_{A}{}^{\ui{M}}$, corresponding to the vacuum we are considering. The mass matrices for the spin-1 and spin-2 Kaluza-Klein modes can be found in \cite{Malek:2019eaz,MalekSamtleben}.\footnote{An $\SL{8}$-covariant mass matrix for the spin-2 sector has been proposed in \cite{Dimmitt:2019qla}}.

Note that not all the fluctuations in $j_{AB,\Sigma}$ are physical. Some of these modes are Goldstone bosons eaten by massive vector fields, and some are eaten by the massive gravitons. We can remove these unphysical modes by fixing the gauge appropriately, as usual when computing Kaluza-Klein spectra.

\subsection{$\SU{8}$ mass matrix}
In the mass matrix \eqref{eq:ScalarMassMatrix}, the $A, B = 1, \ldots, 56$ indices are raised/lowere by $\delta_{AB}$, which is not an invariant tensor of $E_{7(7)}$. Instead, $\delta_{AB}$ breaks $E_{7(7)}$ to its maximal compact subgroup, $\SU{8}$ and, correspondingly, the $A, B$ indices should really be thought of as $\SU{8}$ indices.

Since the dressed embedding tensor $X_{AB}{}^C$ is frequently given in $\SU{8}$-covariant notation, it is worthwhile rewriting the mass matrix \eqref{eq:ScalarMassMatrix} in $\SU{8}$-covariant notation. Let $i = 1, \ldots, 8$ denote fundamental $\SU{8}$ indices and upstairs/downstairs indices be related by complex conjugation. Under the decomposition $E_{7(7)} \rightarrow \SU{8}$, the $\mathbf{56}$ and embedding tensor representation, $\mathbf{912}$, decompose as
\begin{equation}
\begin{split}
\mathbf{56} &\longrightarrow \mathbf{28} \oplus \overline{\mathbf{28}} \,, \\
\mathbf{912} &\longrightarrow \mathbf{36} \oplus \overline{\mathbf{36}} \oplus \mathbf{420} \oplus \overline{\mathbf{420}} \,.
\end{split}
\end{equation}
The $\mathbf{36}$ and $\mathbf{420}$ representations of the embedding tensor are known as the fermion shift matrices, $A_1^{ij}$ and $A_{2,i}{}^{jkl}$. These satisfy
\begin{equation}
 \begin{split}
  A_1^{[ij]} = 0 \,, \qquad A_{2,i}{}^{jkl} = A_{2,i}{}^{[jkl]} \,, \qquad A_{2,i}{}^{ijk} = 0 \,.
 \end{split}
\end{equation}
Explicitly, the relationship between the embedding tensor and the fermion shift matrices is \cite{deWit:2007kvg}
\begin{equation}
\begin{split} \label{eq:EmbTensA}
X_{ij,kl}{}^{pq} &= - X_{ij}{}^{pq}{}_{kl} = \frac12 \delta^{[p}_{[k} A_{2,\phantom{q}l]ij}^{\phantom{2}q]} + \delta^{[p}_{[k} A^{\phantom{[p}}_{1,l][i} \delta_{j]}^{q]} \,, \\
X_{ij}{}^{kl,pq} &= \delta^{[k}_{[i} A^{\phantom{l]}}_{2,j]}{}^{lpq]} \,, \qquad X_{ij,kl,pq} = - \frac1{4!} \epsilon_{klpqtuv[i} A_{2,j]}{}^{tuv} \,,
\end{split}
\end{equation}
with the other components related by complex conjugation to the above.

Similarly, the dressed $\cT_A$ contain the $\SU{8}$ representations $\mathbf{28} \oplus \overline{\mathbf{28}}$ as follows
\begin{equation}
\cT_{A}{}^{\Sigma}{}_{\Omega} = \left( \cT_{ij}{}^{\Sigma}{}_{\Omega} ,\, \cT^{ij,\Sigma}{}_{\Omega} \right) \,,
\end{equation}
with $\cT^{ij,\Sigma}{}_{\Omega}$ the complex conjugate of $\cT_{ij}{}^{\Sigma}{}_{\Omega}$. Finally, since the scalar $j_{AB,\Sigma}$, for fixed $\Sigma$, parameterise the coset $\mathfrak{e}_{7(7)} - \mathfrak{su}(8)$, their only non-zero components are given by
\begin{equation} \label{eq:ScalarSU8}
j^{ijkl}{}_{\Sigma} = \bar{j}_{ijkl,\Sigma} = \frac{1}{4!} \epsilon^{ijklmnpq} j_{mnpq,\Sigma} \,,
\end{equation}
where $\bar{j}_{ijkl,\Sigma}$ is the complex conjugate of $j_{ijkl,\Sigma}$, and $\epsilon^{ijklmnpq}$ is the eight-dimensional alternating symbol.

With the above conventions, the mass matrix \eqref{eq:ScalarMassMatrix} in $\SU{8}$-covariant notation, is given by
\begin{equation}
\begin{split} \label{eq:ScalarMassMatrixSU8}
M^{ijkl\Sigma}{}_{mnpq\Omega}\, j_{ijkl,\Sigma} \, j^{mnpq,\Omega} &= \left( - \frac12 A_1{}^{ij} A_{1,ij} + \frac5{24} A_{2,i}{}^{jkl} A_2{}^i{}_{jkl} \right) j_{klmn,\Sigma} \, j^{klmn,\Sigma} \\
& \quad + \left( 6 A_{2,i}{}^{kmn} A_2{}^{l}{}_{jmn} - \frac32 A_2{}^m{}_{nij} A_{2,m}{}^{nkl} \right) j^{ijpq,\Sigma} \, j_{klpq,\Sigma} \\
& \quad - \frac23 A_{2,i}{}^{mnp} A_2{}^{q}{}_{jkl} j^{ijkl,\Sigma} j_{mnpq,\Sigma} \\
& \quad + 4 \left( A_{2,l}{}^{ijk}\, \cT_{mn,\Omega}{}^{\Sigma} + A_{2}{}^{i}{}_{lmn}\, \cT^{jk,\Sigma}{}_{\Omega} \right) j_{ijkp,\Sigma}\, j^{lmnp,\Omega} \\
& \quad - 4 \left( A_{2,i}{}^{jkl}\, \cT_{kl,\Omega}{}^{\Sigma} + A_2{}^j{}_{ikl}\, \cT^{kl,\Sigma}{}_{\Omega} \right) j_{jmnp,\Sigma}\,  j^{imnp,\Omega} \\
& \quad - 8 \left( A_{1,ik}\, \cT^{kj,\Sigma}{}_{\Omega} + A_1{}^{jk}\, \cT_{ki,\Omega}{}^{\Sigma} \right) j_{jmnp,\Sigma}\,  j^{imnp,\Omega} \\
& \quad + 48\, j_{ijpq,\Sigma}\, j^{klpq,\Omega}\, \cT^{ij}{}_{\Omega}{}^{\Lambda}\, \cT_{kl,\Lambda}{}^{\Sigma} - 2\, j_{ijkl,\Sigma}\, j^{ijkl,\Omega}\, \cT_{mn,\Omega}{}^{\Lambda}\, \cT^{mn}{}_{\Lambda}{}^{\Sigma} \,.
\end{split}
\end{equation}
The first three lines above correspond to the mass matrix of the four-dimensional ${\cal N}=8$ supergravity \cite{deWit:1983gs}, while the last four lines provide corrections of the mass for the higher Kaluza-Klein levels.

\section{The $\SO{3} \times \SO{3}$-invariant AdS$_4$ vacuum} \label{s:Vacuum}
The $\SO{3} \times \SO{3}$-invariant AdS$_4$ vacuum was first found as a solution of four-dimensional ${\cal N}=8$ $\SO{8}$-gauged SUGRA \cite{Warner:1983du,Warner:1983vz}. Since the $\SO{8}$-gauged SUGRA arises as a consistent truncation of 11-dimensional supergravity on $S^7$ \cite{deWit:1986oxb}, this non-supersymmetric AdS$_4$ vacuum can be uplifted to a solution of 11-dimensional supergravity. Concretely, this is done by making use of the known uplift formulas for the internal metric \cite{deWit:1984nz} and the internal components of the three-form \cite{deWit:2013ija}. For the  $\SO{3} \times \SO{3}$-invariant vacuum the resulting 11-dimensional solution has been worked out in \cite{Godazgar:2014eza}. As the relevant expressions are rather complicated and not needed here, we refer readers there for further details and explicit formulas.

Indeed, using the methods of \cite{Malek:2019eaz,MalekSamtleben}, we only need the explicit form of the four-dimensional scalar matrix at the $\SO{3} \times \SO{3}$ stationary point \cite{Warner:1983du} in order to compute the Kaluza-Klein spectrum. In four-dimensional ${\cal N}=8$ supergravity, the scalar matrix is an $E_{7(7)}/\SU{8}$ coset element, $\mathcal{V}_A{}^M$, in the fundamental representation \textbf{56} of E$_{7(7)}$ \cite{Cremmer:1978ds}
\begin{equation}
\mathcal{V}_A{}^M(x) = \begin{pmatrix}
u_{ij}{}^{IJ} (x) & v^{ijIJ} (x)\\ 
v_{ijIJ}(x) & u^{ij}{}_{IJ}(x)
\end{pmatrix}\,.
\end{equation}
Here $i,j =1, \ldots, 8$ denote $\SU{8}$ indices, while $I,J = 1, \ldots, 8$ denote $\SL{8}$ indices; complex conjugation is realised by raising/lowering of all indices, for example $u^{ij}{}_{IJ} = (u_{ij}{}^{IJ})^*$. By exploiting the local SU(8) invariance we can bring $\mathcal{V}$ into a unitary gauge, where it takes the form
\begin{equation}\label{eq:56bein}
\mathcal{V} = \exp
\begin{pmatrix}
0 & \phi_{IJKL} \\ \phi^{IJKL} & 0     
\end{pmatrix}  \,\equiv\,
\begin{pmatrix}
u_{IJ}{}^{KL} & v^{IJKL} \\ 
v_{IJKL} & u^{IJ}{}_{KL}
\end{pmatrix}\,,
\end{equation}
where after gauge fixing we no longer need to distinguish between $\SO{8}$ and $\SU{8}$ indices. In this gauge, the $u$ and $v$ matrices are expressed as infinite sums, {\it viz.}
\begin{equation}
u_{IJ}{}^{KL} = \sum_{n=0}^\infty \frac{1}{(2n)!}  \left[ (\phi\phi^*)^n  \right]_{IJKL}\,, 
\qquad v^{IJKL} = \sum_{n=0}^\infty \frac{1}{(2n+1)!} \left[ \phi^* (\phi\phi^*)^n \right]_{IJKL}\,.
\end{equation}

To give the scalar matrix at the $\SO{3} \times \SO{3}$ invariant point, we introduce the four $\SO{3} \times \SO{3}$-invariant tensors
\begin{equation}
 \begin{split}
  Y^+_{IJKL} &= 4!  \left(\delta^{1234}_{IJKL} + \delta^{5678}_{IJKL} \right)\,, \qquad
  Y^-_{IJKL} = 4!  \left(\delta^{1235}_{IJKL} + \delta^{4678}_{IJKL} \right)\,, \\
  Z^-_{IJKL} &= 4!  \left(\delta^{1234}_{IJKL} - \delta^{5678}_{IJKL} \right)\,, \qquad
  Z^+_{IJKL} = 4!  \left(\delta^{1235}_{IJKL} - \delta^{4678}_{IJKL} \right)\,, \label{eq:so8tensors}
 \end{split}
\end{equation}
where the two SO(3) subgroups act on the subspaces defined by $I=1,2,3$ and $I = 6,7,8$, respectively.  Note that tensors $Y^+$ and $Z^+$ are self-dual, while $Y^-$ and $Z^-$  are antiself-dual (another $\SO{3} \times \SO{3}$ invariant is $F_{IJ} = \delta^{45}_{IJ}$). These objects satisfy a number of elementary identities which have been listed in \cite{Godazgar:2014eza}.

With these definitions, we can parametrize the SO(3) $\times$ SO(3) invariant scalar field configurations as
\begin{equation}
\phi_{IJKL}(\lambda,\omega)  = \frac{\lambda}{2} \Big[ \cos \omega \left( Y^+_{IJKL} + i Y^-_{IJKL} \right) 
- \sin \omega \left( Z^+_{IJKL} - i Z^-_{IJKL} \right) \Big] \,,
\end{equation}
with two independent parameters $\lambda$ and $\omega$. To exponentiate the scalar expectation value it is, furthermore, useful to define the Hermitian projector~\footnote{With the short-hand notation 
	$A \, B \equiv  ( A \, B )_{IJKL}  \equiv  A_{IJMN} B_{MNKL}$.}
\begin{equation}
\label{Pi}
\Pi = \frac{1}{8} \left( Y^+ + i Y^- \right) \left( Y^+ - i Y^- \right) = \frac{1}{8} \left( Z^+ -i Z^- \right) \left( Z^+ + i Z^- \right)\,,
\end{equation}
which satisfies 
\begin{equation}
\Pi^2 = \Pi\,, \quad \Pi^*_{IJKL} = \Pi_{KLIJ}\,, \quad \left( Y^+ - i Y^- \right) \Pi = Y^+ - i Y^-, \quad \left( Z^+ + i Z^- \right)
\Pi = Z^+ + i Z^-.
\end{equation}
so $\Pi$ is a Hermitian projector. In particular, using identities from \cite{Godazgar:2014eza}, we find that
\begin{equation}
\phi \phi^* = 2 \lambda^2 \Pi\,, \qquad \phi^* \Pi = \phi^*\,.
\end{equation}
After these preparations it is straightforward to determine the $u$ and $v$ matrices 
\begin{equation} \label{eq:ScMSO3SO3}
 \begin{split}
  u_{IJ}{}^{KL} &= \delta_{IJ}^{KL} + \big( \cosh(\sqrt{2}\lambda) - 1\big) \Pi_{IJKL}\,, \\
  v^{IJKL} &= \frac{1}{2\sqrt{2}} \sinh(\sqrt{2}\lambda) \Big[ \cos \omega \, (Y^+ - i Y^-) - \sin \omega  \,(Z^+ + i Z^-) \Big]_{IJKL}\,,
 \end{split}
\end{equation}

With these explicit expressions for the scalar matrix \eqref{eq:ScMSO3SO3}, we can now compute the dressed embedding tensor, or equivalently via \eqref{eq:EmbTensA} the fermion shift matrices, $A_{1}^{ij}$ and $A_{2\,i}{}^{jkl}$, for any value of $\lambda$ and $\omega$, which includes the $\SO{3} \times \SO{3}$-invariant vacuum. For the $\SO{8}$-gauged supergravity we are considering, the fermion shift matrices can be computed via
\begin{equation}
(u^{kl}{}_{IJ}  + v^{klIJ})(u_{im}{}^{JK} u^{jM}{}_{KI} - v_{imJK} v^{jmKI}) \,= \,
-\frac34A_{2\,i}{}^{jkl} + \frac32 \delta_i^{[k} A_1^{l]j} \,.
\end{equation}
For the above scalar field configuration, we get \cite{Warner:1983du,Bobev:2011rv}
\begin{equation}  \label{eq:A1}
A_1^{ij}(\lambda) = {\rm diag}\,\big( a,a,a,b,b,a,a,a\big) \,,
\end{equation}
with
\begin{equation}
a(\lambda) = \cosh(\sqrt{2}\lambda) \,, \qquad 
b(\lambda) = \frac12 \left(1+\cosh^2(\sqrt{2}\lambda)\right) \,.
\end{equation}
For $A_{2\,i}{}^{jkl}(\lambda)$ we have \cite{Bobev:2011rv}
\begin{equation}
 \begin{split} \label{eq:A2}
  (A_2)_1{}^{234} & = (A_2)_2{}^{314}  = (A_2)_3{}^{124}  = - e^{-i\omega} f \,, \\
(A_2)_1{}^{235} & = (A_2)_2{}^{315}  = (A_2)_3{}^{125}  = i e^{-i\omega} f \,, \\
(A_2)_6{}^{784} & = (A_2)_7{}^{864}  = (A_2)_8{}^{674}  = - i e^{i\omega} f \,, \\
(A_2)_6{}^{785} & = (A_2)_7{}^{865}  = (A_2)_8{}^{675}  = e^{i\omega} f \,, \\
(A_2)_4{}^{678} & = ie^{i\omega} g \;,\quad  (A_2)_5{}^{123} = - i e^{- i\omega} g \,, \\
(A_2)_4{}^{123} & = e^{-i\omega} g \; ,\quad  (A_2)_5{}^{678}  = - e^{i\omega}  g \,, \\
(A_2)_1{}^{145} &=  (A_2)_2{}^{245} = (A_2)_3{}^{345}  =
- (A_2)_6{}^{645} = - (A_2)_7{}^{745}  = - (A_2)_8{}^{845}  = - i f^2 \,,
 \end{split}
\end{equation}
with
\begin{equation}
f(\lambda) = \frac1{\sqrt{2}}\sinh(\sqrt{2}\lambda) \,, \qquad 
g (\lambda) = \frac1{2\sqrt{2}}\sinh(2\sqrt{2}\lambda) \,.
\end{equation}

The potential of the four-dimensional supergravity depends only on $A_1^{ij}$ and $A_{2\,i}{}^{jkl}$ and for the field configurations considered here is given by
\begin{equation}
 \begin{split}
{\cal P}(\lambda,\omega) & = - \frac34(6a^2 +2b^2) + \frac14(12 f^2 + 6 f^4 + 4 g^2) \\
& = \frac1{16}\Big[ - 36 \cosh(2\sqrt{2}\lambda) + \cosh(4\sqrt{2}\lambda) - 61\Big] \,,
\end{split}
\end{equation}
and does not depend on $\omega$. As shown in \cite{Godazgar:2014eza}, a rotation by an angle $\omega$ corresponds to a diffeomorphism in the internal dimensions, hence does not change the physical solution. The extremum is attained at
\begin{equation}
\lambda = \frac1{\sqrt{2}} {\rm arcosh} ( \sqrt{5}) \quad
\Leftrightarrow \quad \cosh(\sqrt{2}\lambda) = \sqrt{5} \,,
\end{equation}
so that
\begin{equation}
a = \sqrt{5}\,,\quad b=3 \,, \qquad
f = \sqrt{2} \,, \quad g = \sqrt{10} \,,
\end{equation}
at the $\SO{3} \times \SO{3}$ stationary point, with ${\mathcal{P}} = -14$. The point $\lambda = 0$ corresponds to the maximally supersymmetric AdS$_4$ vacuum with $\SO{8}$ symmetry.

As mentioned previously, this stationary point is the only known stable non-supersymmetric stationary point of $D=4$ $\SO{8}$ gauged supergravity \cite{Warner:1983du,Fischbacher:2010ec}: the scalar mass eigenvalues for all 48 physical scalar fields are above the BF bound (22 scalars are Goldstone bosons, and thus are absorbed by the 22 massive gauge fields remaining after symmetry breaking to $\SO{3} \times \SO{3}$). Since for maximal AdS supergravities in other dimensions $D > 4$ no stable non-supersymmetric AdS vacua are known, this vacuum has been one of the most promising candidates for a stable non-supersymmetric AdS$_4$ solution of M theory. The critical question is, therefore, whether the Kaluza-Klein spectrum can produce a tachyonic instability, and this is the question that will be addressed in the following section. We note that by the mass formula \eqref{eq:ScalarMassMatrixSU8}, we only require the knowledge of $A_1$ and $A_2$ at the stationary point \eqref{eq:A1} and \eqref{eq:A2}, and the dressed action of Killing vectors on $S^7$ to compute the Kaluza-Klein spectrum.

\section{Scalar harmonics} \label{s:Harmonics}
To compute the Kaluza-Klein spectrum, we now only need the higher-dimensional information coming from the round $S^7$. The scalar harmonics on $S^7$ can be expressed as symmetric traceless polynomials in the elementary harmonics ${\cal Y}^a$, with $a = 1, \ldots, 8$, which satisfy ${\cal Y}^a {\cal Y}^b\, \delta_{ab} = 1$. Thus,
\begin{equation}
\left\{ {\cal Y}^\Sigma \right\} = \left\{ 1,\, {\cal Y}^a,\, {\cal Y}^{a_1a_2},\, \ldots ,\, {\cal Y}^{a_1 \ldots a_n},\, \ldots \right\} \,,
\end{equation}
where ${\cal Y}^{a_1 \ldots a_n} \equiv {\cal Y}^{(\!(a_1} \ldots {\cal Y}^{a_n)\!)}$ denotes traceless symmetrisation. Hence the index $\Sigma$ runs over the tower of symmetric traceless representations $\left[ n,\, 0,\, 0,\, 0 \right]$ of $\SO{8}$.

The action of the $\SO{8}$ Killing vectors of the round $S^7$ on the harmonics is most easily described in the $\SL{8} \subset E_{7(7)}$ basis. Under this $\SL{8}$, the $\mathbf{56}$ of $E_{7(7)}$ decomposes as
\begin{equation}
\mathbf{56} \longrightarrow \mathbf{28} \oplus \mathbf{28}' \,.
\end{equation}
Accordingly, the action of the $\SO{8}$ Killing vectors decomposes into
\begin{equation}
\begin{split}
\cT_{M}{}^{c_1 \ldots c_n}{}_{d_1 \ldots d_m} &= \left( \cT_{ab},\, \cT^{ab} \right)^{c_1 \ldots c_n}{}_{d_1 \ldots d_m} \,.
\end{split}
\end{equation}
Our summation convention for the harmonic indices $\Sigma, \Omega$ is such that
\begin{equation}
A^{\Sigma}\, B_{\Sigma} = A\,B + A^a\, B_a + A^{a_1a_2}\, B_{a_1a_2}\, + \ldots + A^{a_1\ldots a_n}\, B_{a_1 \ldots a_n} + \ldots
\end{equation}
With this convention, we find
\begin{equation}
\begin{split}
\cT_{ab}{}^{c_1 \ldots c_n}{}_{d_1 \ldots d_n} &= c\, n\, \delta^{(\!(c_1}_{[a} \delta^{\phantom{(\!(a}}_{b](\!(d_1} \delta^{c_2}_{d_2} \ldots \delta^{c_n)\!)}_{d_n)\!)} \,, \\
\cT^{ab,c_1 \ldots c_n}{}_{d_1 \ldots d_n} &= 0 \,,
\end{split}
\end{equation}
where $c$ is a coefficient that is determined by the normalisation \eqref{eq:TNorm}, i.e.
\begin{equation}
 \left[ \cT_{M},\, \cT_{N} \right] = X_{[MN]}{}^P\, \cT_P \,.
\end{equation}

In order to match the notation of section \ref{s:Vacuum}, we must convert between the $\SL{8}$ basis used here and the $\SU{8}$ basis used there. We explain how to do this in appendix \ref{app:SU8SL8}. In the conventions of section \ref{s:Vacuum}, the round $S^7$ has $A_{1}^{ij} = \delta^{ij}$ and $A_{2\,i}{}^{jkl} = 0$. Therefore, the corresponding embedding tensor in the $\SL{8}$ basis is given by
\begin{equation}
 \begin{split}
  X_{ab\,cd}{}^{ef} &= - X_{ab}{}^{ef}{}_{cd} = - \frac{1}{\sqrt{2}} \left( \delta_{b[c}^{\phantom{ef}} \delta_{d]a}^{ef} - \delta_{a[c}^{\phantom{ef}} \delta_{d]b}^{ef} \right) \,, \\
  X^{ab\,cd\,ef} &= X_{ab}{}^{cd\,ef} = 0 \,, \\
  X^{ab}{}_{M}{}^N &= 0 \,,
 \end{split}
\end{equation}
and the normalisation coefficient for the $\cT_M$ is $ c = \frac{1}{\sqrt{2}}$. Thus, the only non-zero components of the $\cT_M$ are
\begin{equation}
 \cT_{ab}{}^{c_1 \ldots c_n}{}_{d_1 \ldots d_n} = \frac{n}{\sqrt{2}} \delta^{(\!(c_1}_{[a} \delta^{\phantom{(\!(a}}_{b](\!(d_1} \delta^{c_2}_{d_2} \ldots \delta^{c_n)\!)}_{d_n)\!)} \,.
\end{equation}
The action of the $\SO{8}$ Killing vectors, dressed by the scalar matrix of the $\SO{3}\times\SO{3}$-invariant AdS vacuum \eqref{eq:ScMSO3SO3}, (in the $\SU{8}$ basis) is now given by
\begin{equation}
 \begin{split} \label{eq:DressedT}
  \cT_{IJ}{}^{c_1 \ldots c_n}{}_{d_1\ldots d_n} &= \frac{1}{4\sqrt{2}} \left[ \Gamma^{ab}_{IJ} + \left( \cosh(\sqrt{2}\lambda) - 1 \right) \Pi_{IJKL} \Gamma^{ab,KL} \right. \\
  & \quad \left. + \frac{1}{2\sqrt{2}} \sinh(\sqrt{2}\lambda) \left( \cos\omega \left(Y^+ + i Y^-\right) - \sin\omega \left(Z^+ - iZ^-\right)\right)_{IJKL} \Gamma^{ab,KL} \right] \cT_{ab}{}^{c_1 \ldots c_n}{}_{d_1\ldots d_n} \,,
 \end{split}
\end{equation}
where $\Gamma^{ab}{}_{IJ}$ are the $\Gamma$-matrices, see appendix \ref{app:SU8SL8}.

\section{The Kaluza-Klein spectrum} \label{s:Spectrum}
We will now apply the method of \cite{Malek:2019eaz,MalekSamtleben}, reviewed in section \ref{s:Spectroscopy}, to the $\SO{3} \times \SO{3}$-invariant AdS$_4$ solution \cite{Warner:1983du,Warner:1983vz} discussed in section \ref{s:Vacuum}. We have already discussed the fermion shift matrices $A_1^{ij}$ and $A_{2\,i}{}^{jkl}$ at the $\SO{3} \times \SO{3}$-invariant point, eqs. \eqref{eq:A1}, \eqref{eq:A2}, and the action of the dressed Killing vectors on the scalar harmonics $\cT_{A}{}^{\Sigma}{}_{\Omega}$, eq. \eqref{eq:DressedT}, which we require to compute the Kaluza-Klein spectrum.

The Kaluza-Klein modes organise themselves into representations of the $\SO{3} \times \SO{3}$ symmetry group. This is embedded into the $\SO{8}$ symmetry group of the round $S^7$ according to the following branching of $\SO{8} \longrightarrow \SO{3} \times \SO{3}$:
\begin{equation}
 \begin{split}
  \mathbf{8}_c &\longrightarrow \mathbf{\left(3,1\right)} \oplus \mathbf{\left(1,3\right)} \oplus 2 \cdot \mathbf{\left(1,1\right)} \,, \\
  \mathbf{8}_{v,s} &\longrightarrow 2 \cdot \mathbf{\left(2,2\right)} \,,
 \end{split}
\end{equation}
where the Kaluza-Klein scalar harmonics transform in the $\mathbf{8}_v$.

Using this branching, we can determine the $\SO{3} \times \SO{3}$ representation of the Kaluza-Klein modes at level $n$. We do this by taking the tensor product of the ${\cal N}=8$ supergravity modes with the scalar harmonics at level $n$ and decompose the result under $\SO{3} \times \SO{3}$. Indeed, the Kaluza-Klein fluctuation Ansatz of \cite{Malek:2019eaz,MalekSamtleben} reflects this group-theoretic analysis at the level of the linearised dynamics by writing the fluctuation Ansatz as a product of the consistent truncation (which encapsulates the ${\cal N}=8$ supergravity modes) with the scalar harmonics.

For example, for the spin-2 Kaluza-Klein modes, we simply need to branch the $\left[n,0,0,0\right]$ representation of $\SO{8}$, in which the scalar harmonics transform, to $\SO{3} \times \SO{3}$ and find
\begin{equation}
\begin{split}
G_{n} &= \bigoplus_{l=0}^{n/2} \left(n-2l+1\right)\left(l+1\right) \cdot \mbf{n-2l+1}{n-2l+1} \\
& \quad \oplus \bigoplus_{l=1}^{n/2} \bigoplus_{k=0}^{l-1} \left(n-2l+1\right)\left(k+1\right) \cdot \Big[ \mbf{n-2l+1}{n-2k+1} \oplus \mbf{n-2k+1}{n-2l+1} \Big] \,.
\end{split}
\end{equation}
Similarly, we can determine the representations of the spin-1 Kaluza-Klein modes, which we denote by $\tilde{V}_n$ at level $n$, by taking the tensor product of the $\mathbf{28}$ of $\SO{8}$, corresponding to the ${\cal N}=8$ vectors, with the $\left[n,0,0,0\right]$ representation of $\SO{8}$. Finally, for the spin-0 KK modes, which we denote by $\tilde{S}_n$ at level $n$, we take the tensor product of the $\mathbf{35}_v \oplus \mathbf{35}_s$ of $\SO{8}$, corresponding to the ${\cal N}=8$ scalars, with the $\left[n,0,0,0\right]$ representation of $\SO{8}$. However, the resulting tensor products, just like the fluctuation Ansatz \eqref{eq:ScalarFluct}, contain also unphysical modes, in particular Goldstone scalars and vectors which are eaten by the massive vector fields and massive Gravitons. These can be removed by subtracting the massive Graviton representations from the massive vector representations and the massive vector representations, $\tilde{V}_n$, from the massive scalar representations. We denote the resulting physical Kaluza-Klein modes at level $n$ by $V_n$ and $S_n$ and find
\begin{equation}
\begin{split}
 V_{n} &= \left(n+1\right) \cdot \mbf{n+3}{n+3} \oplus \left(5\,n+3\right) \cdot \left[ \mbf{n+3}{n+1} \oplus \mbf{n+1}{n+3} \right]  \\
 & \quad \oplus \bigoplus_{l=0}^{(n-1)/2} \left( \left(27\,l-1\right) \left(n-2\,l\right) + 18\,n - 29\,l - 3 \right) \cdot \mbf{n+1-2l}{n+1-2l} \\
 & \quad \oplus \bigoplus_{l=1}^{n/2} 6 \left(n+1-2\,l\right) \cdot \left[ \mbf{n+3}{n+1-2l} \oplus \mbf{n+1-2l}{n+3} \right] \\
 & \quad \oplus \bigoplus_{l=1}^{n/2} \left( 27\,l \left(n+1-2\,l\right) - \left(n+3\right) \right) \cdot \left[ \mbf{n+3-2l}{n+1-2l} \oplus \mbf{n+1-2l}{n+3-2l} \right] \\
 & \quad \oplus \bigoplus_{l=2}^{n/2} \bigoplus_{k=0}^{l-2} 27 \left(k+1\right) \left(n+1-2\,l\right) \cdot \left[ \mbf{n+1-2k}{n+1-2l} \oplus \mbf{n+1-2l}{n+1-2k} \right] \\
 & \quad \oplus \frac{9\,n}4 \left(1+\left(-1\right)^n\right) \cdot \mbf{1}{1} \,,
\end{split}
\end{equation}
\begin{equation}
 \begin{split}
 S_{n} &= 5 \left(n+1\right) \cdot \mbf{n+3}{n+3} \oplus \left(9\,n-1\right) \cdot \left[ \mbf{n+3}{n+1} \oplus \mbf{n+1}{n+3} \right]  \\
 & \quad \oplus \bigoplus_{l=0}^{(n-1)/2} 2 \left( 21\,l \left(n-1-2\,l\right) + 12\,n - 6 \right) \cdot \mbf{n+1-2l}{n+1-2l} \\
 & \quad \oplus \bigoplus_{l=1}^{n/2} 14 \left(n+1-2\,l\right) \cdot \left[ \mbf{n+3}{n+1-2l} \oplus \mbf{n+1-2l}{n+3} \right] \\
 & \quad \oplus \bigoplus_{l=1}^{n/2} \left( 42\,l \left(n+1-2\,l\right) - 5\left(n+3\right) \right) \cdot \left[ \mbf{n+3-2l}{n+1-2l} \oplus \mbf{n+1-2l}{n+3-2l} \right] \\
 & \quad \oplus \bigoplus_{l=2}^{n/2} \bigoplus_{k=0}^{l-2} 42 \left(k+1\right) \left(n+1-2\,l\right) \cdot \left[ \mbf{n+1-2k}{n+1-2l} \oplus \mbf{n+1-2l}{n+1-2k} \right] \\
 & \quad \oplus \frac12 \left(1+\left(-1\right)^n\right) \left(3+8\,n\right) \cdot \mbf{1}{1} \,.
\end{split} \label{eq:ScalarReps}
\end{equation}

\begin{figure}
	\centering
	\includegraphics[width=\textwidth]{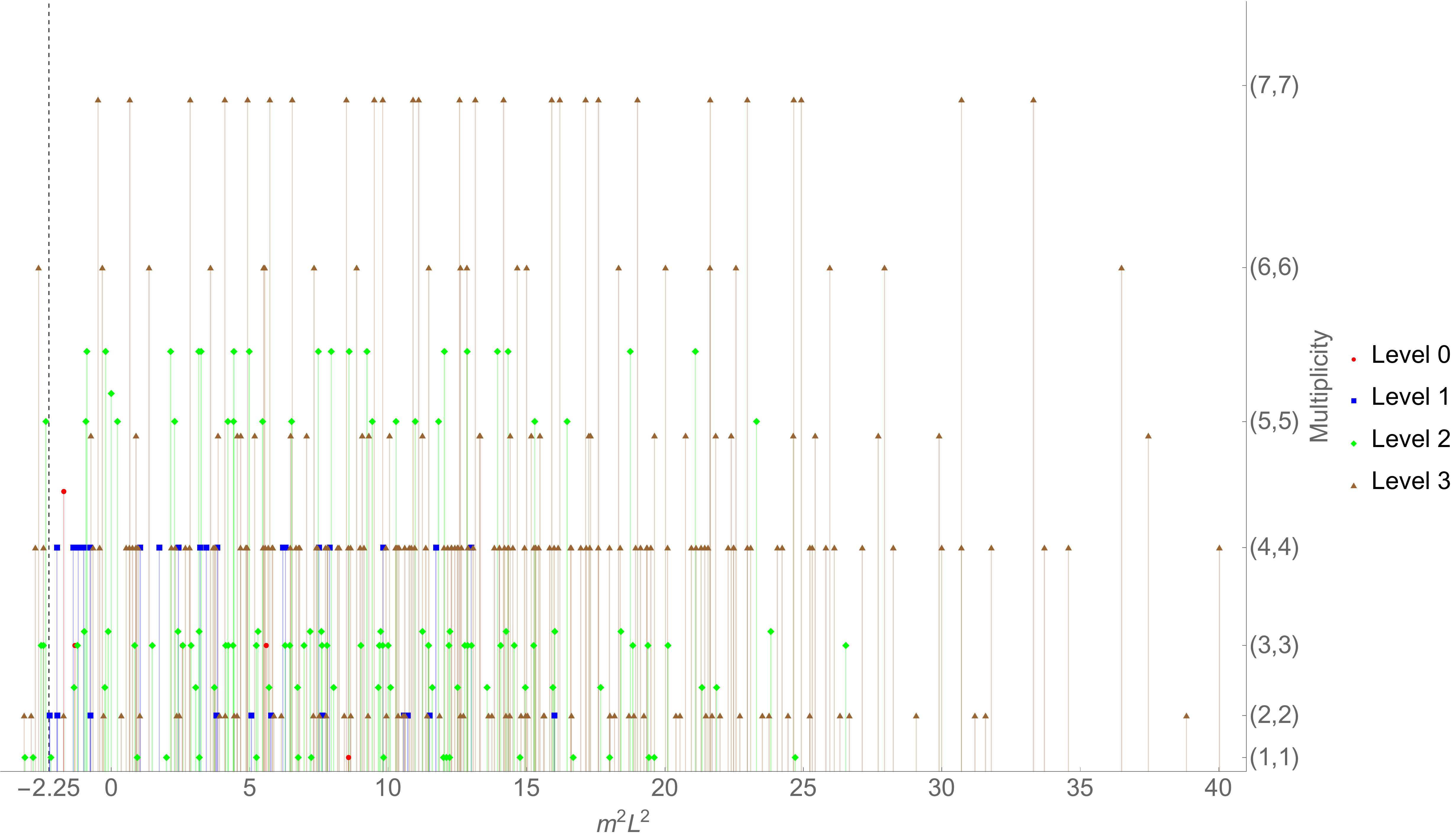}\\[5mm]
	\includegraphics[width=\textwidth]{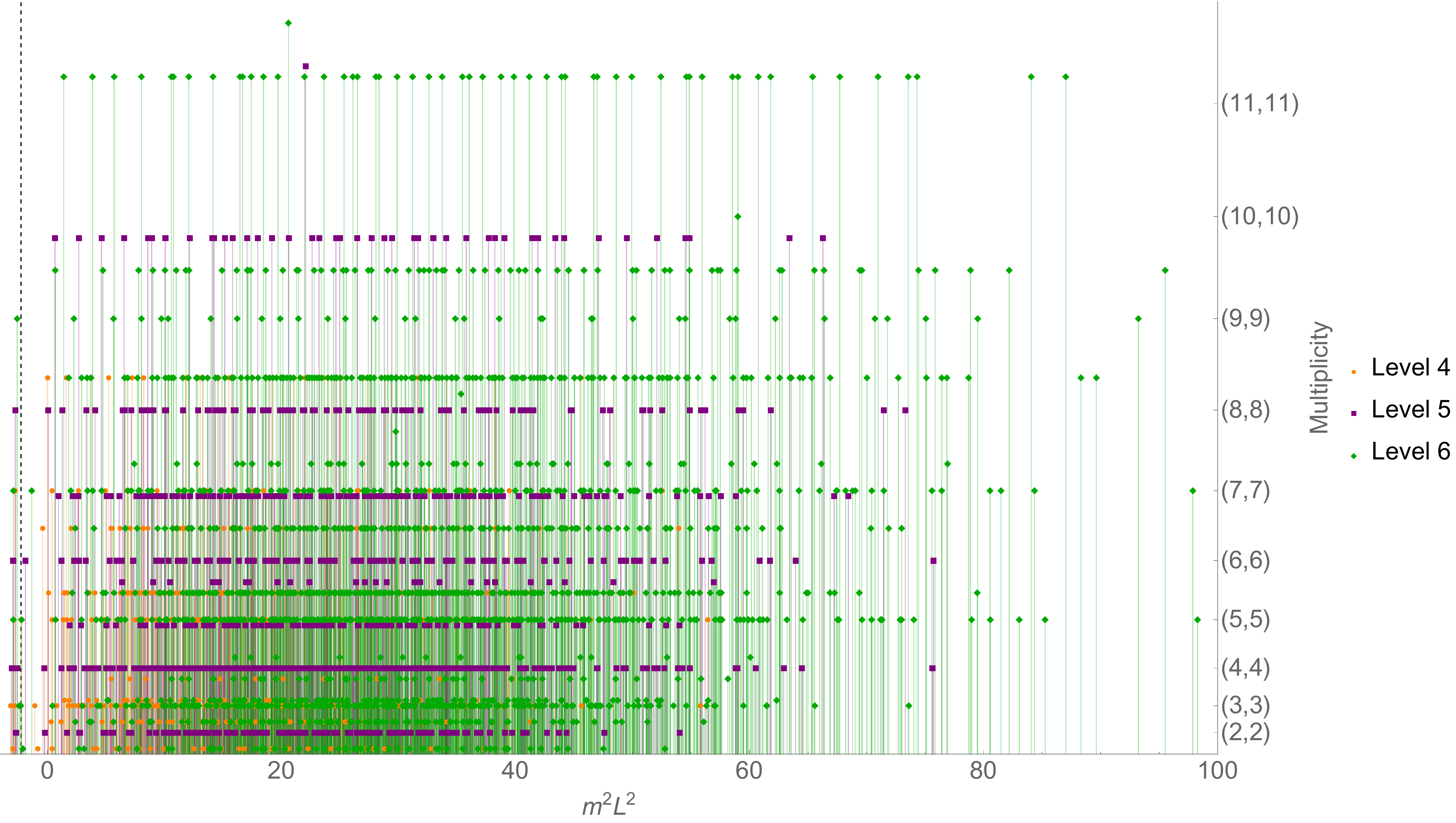}
	\captionof{figure}{\small Spectrum of scalar Kaluza-Klein modes, plotted as the multiplicity of Kaluza-Klein mass eigenvalues for levels  0 -- 6. $m$ is the mass of the Kaluza-Klein mode, $L$ is the AdS length and the multiplicity is determined by the $\SO{3} \times \SO{3}$ representation, $\mbf{k}{k}$, of the mode. The dashed line denotes the BF bound $m^2 L^2 = - 2.25$.} \label{f:Spectrum}
\end{figure}

\begin{figure}
	\centering
	\includegraphics[width=\textwidth]{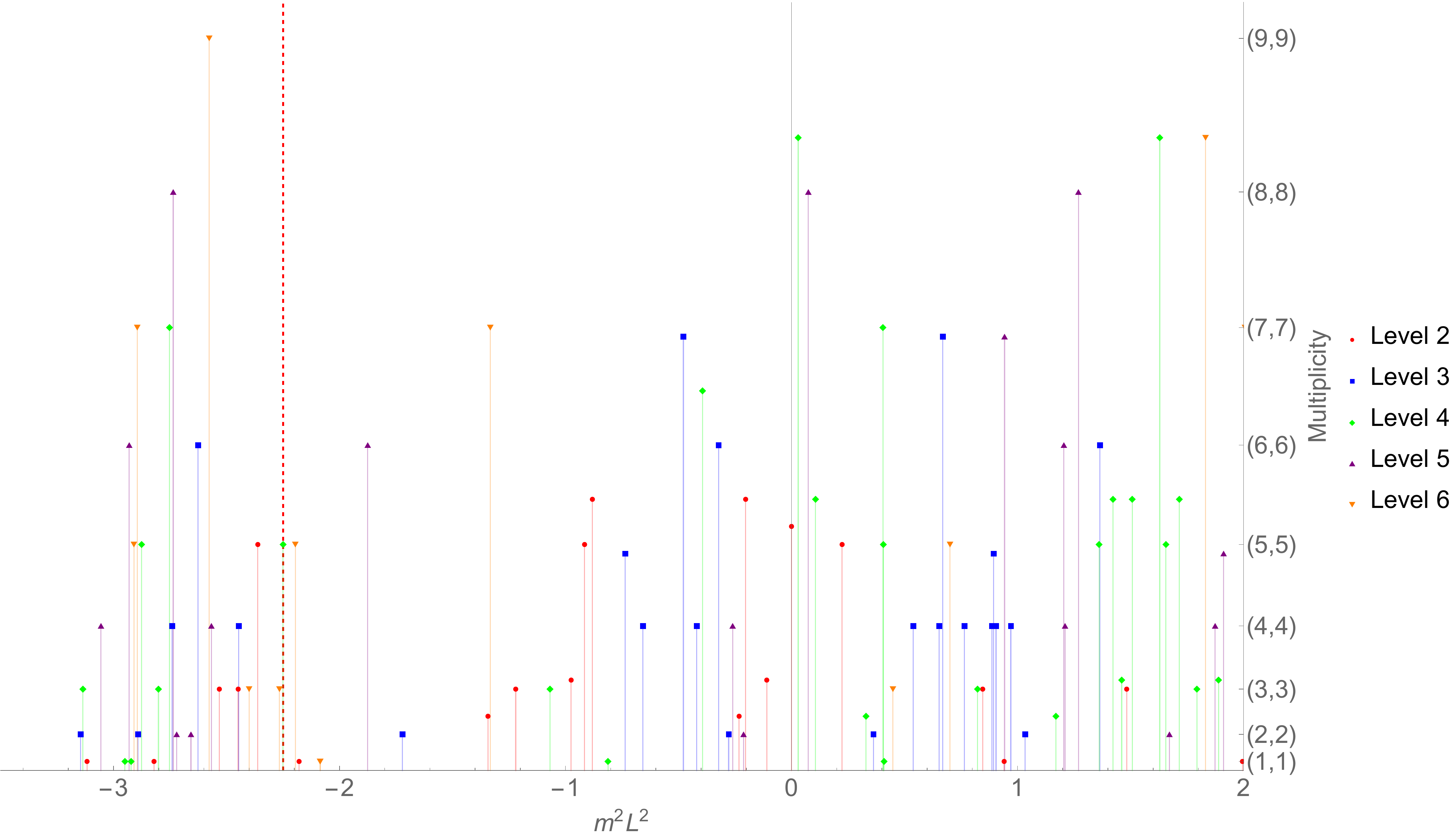}
	\captionof{figure}{\small Spectrum of scalar Kaluza-Klein modes with masses in the range $-3.5 \leq m^2 L^2 \leq 2$ with the BF bound $m^2 L^2 = -2.25$ represented as a dashed red line. Here, $m$ is the mass of the mode, $L$ is the AdS length and the vertical axis shows the multiplicity which is determined by the $\SO{3} \times \SO{3}$ representation of the modes.}\label{f:Tachs}
\end{figure}

\begin{figure}
	\centering
	\includegraphics[width=\textwidth]{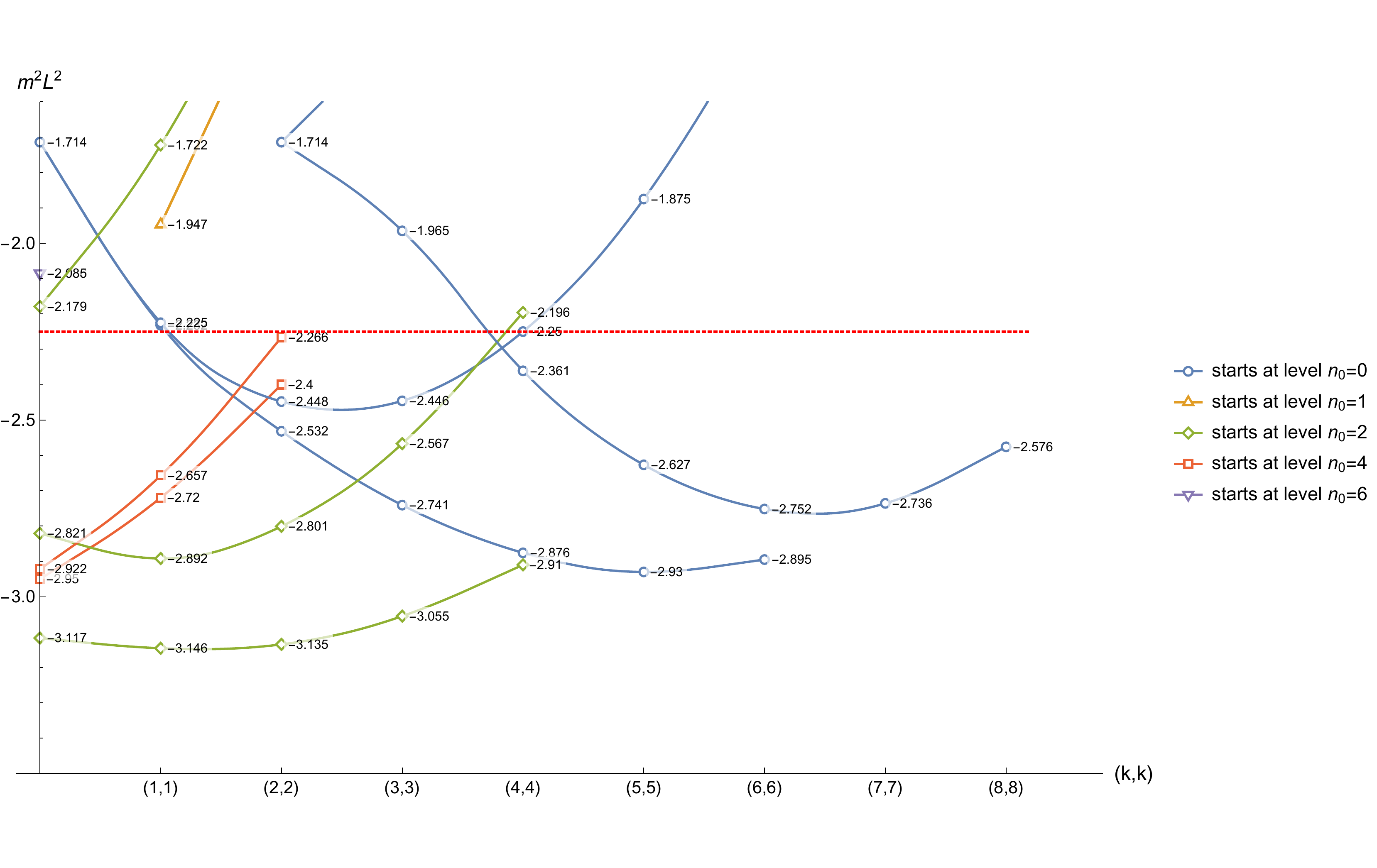}
	\captionof{figure}{\small Lowest Kaluza-Klein modes in the representations $\mbf{k}{k}$ of $\SO{3} \times \SO{3}$ arranged into Kaluza-Klein towers, such that the $n$th dot on a line starting at level $n_0$ with $k = k_0$ sits at level $n_0 + n$ with $k = k_0 + n$. The colour code indicates the starting level $n_0$ for the various towers. The dashed red line corresponds to the BF bound.} \label{f:Towers}
\end{figure}

We can now compute the masses of these Kaluza-Klein modes using the mass matrix \eqref{eq:ScalarMassMatrix}. We use \emph{Mathematica} to compute the eigenvalues of this matrix and trust in its capacities for handling large numerical matrices. Projecting the mass matrices onto the different irreducible $\SO{3} \times \SO{3}$ in \eqref{eq:ScalarReps}, we were able to compute the eigenvalues up to and including level 6 above the four-dimensional ${\cal N}=8$ supergravity modes with sufficient confidence. As a first consistency check the numerical results precisely reproduce the correct massless Goldstone modes. As another check, we work with traceful harmonics ${\cal Y}^{(a_1} \dots {\cal Y}^{a_n)}$, such that the computation at $n=6$ includes and reproduces all previous even levels. Our results are plotted in figures \ref{f:Spectrum} -- \ref{f:Towers}.

The Kaluza-Klein spectrum shows two interesting features: firstly, there are tachyonic modes amongst the higher Kaluza-Klein modes, and secondly, there are 27 infinitesimal moduli, i.e. massless scalar fields which are not eaten by the massive vector fields, at level 2.

\paragraph{Tachyonic Kaluza-Klein modes} The level 0 scalar fields, $S_0$, i.e. those of ${\cal N}=8$ supergravity, have masses at and above the BF bound \cite{Fischbacher:2010ec}. We find this is also true for $S_1$, the scalars at level 1. However, starting at level 2 above the ${\cal N}=8$ scalars, there are \emph{tachyonic} scalar fields amongst the Kaluza-Klein modes, i.e. some spin-0 Kaluza-Klein modes have masses \emph{below} the BF bound, corresponding to $m_{BF}^2 = -2.25 L^{-2}$, where $L$ is the AdS length. Therefore, the AdS$_4$ vacuum is perturbatively unstable within 11-dimensional supergravity. We list the tachyonic modes and their mass eigenvalues in table \ref{t:Tach}.
\begin{figure}[h]
	\begin{center}
		\begin{tabular}{|c|c|}
			\hline
			\multicolumn{2}{|c|}{Level 2} \\
			\hline
			$m^2 L^2$ & Irrep \\
			\hline
			-3.117 & $\mbf{1}{1}$ \\
			-2.821 & $\mbf{1}{1}$ \\
			-2.532 & $\mbf{3}{3}$ \\
			-2.448 & $\mbf{3}{3}$ \\
			-2.361 & $\mbf{5}{5}$ \\
			\hline
		\end{tabular}
		\quad
		\begin{tabular}{|c|c|}
			\hline
			\multicolumn{2}{|c|}{Level 3} \\
			\hline
			$m^2 L^2$ & Irrep \\
			\hline
			-3.146 & $\mbf{2}{2}$ \\
			-2.892 & $\mbf{2}{2}$ \\
			-2.741 & $\mbf{4}{4}$ \\
			-2.446 & $\mbf{4}{4}$ \\
			-2.627 & $\mbf{6}{6}$ \\
			\hline
		\end{tabular}
		\quad
		\begin{tabular}{|c|c|}
			\hline
			\multicolumn{2}{|c|}{Level 4} \\
			\hline
			$m^2 L^2$ & Irrep \\
			\hline
			-2.950 & $\mbf{1}{1}$ \\
			-2.922 & $\mbf{1}{1}$ \\
			-3.114 & $\mbf{3}{3}$ \\
			-2.801 & $\mbf{3}{3}$ \\
			-2.876 & $\mbf{5}{5}$ \\
			-2.752 & $\mbf{7}{7}$ \\
			\hline
		\end{tabular}
		\quad
		\begin{tabular}{|c|c|}
			\hline
			\multicolumn{2}{|c|}{Level 5} \\
			\hline
			$m^2 L^2$ & Irrep \\
			\hline
			-2.721 & $\mbf{2}{2}$ \\
			-2.657 & $\mbf{2}{2}$ \\
			-3.056 & $\mbf{4}{4}$ \\
			-2.567 & $\mbf{4}{4}$ \\
			-2.930 & $\mbf{6}{6}$ \\
			-2.736 & $\mbf{8}{8}$ \\
			\hline
		\end{tabular}
		\quad
		\begin{tabular}{|c|c|}
			\hline
			\multicolumn{2}{|c|}{Level 6} \\
			\hline
			$m^2 L^2$ & Irrep \\
			\hline
			-2.400 & $\mbf{3}{3}$ \\
			-2.266 & $\mbf{3}{3}$ \\
			-2.910 & $\mbf{5}{5}$ \\
			-2.895 & $\mbf{7}{7}$ \\
			-2.577 & $\mbf{9}{9}$ \\
			\hline			
		\end{tabular}
		\captionof{table}{\small Tachyonic Kaluza-Klein modes, their masses, $m$, in terms of the AdS length $L$ and their representations under $\SO{3} \times \SO{3}$.} \label{t:Tach}
	\end{center}
\end{figure}

Interestingly, the tachyonic modes only appear in the symmetric representations $\mbf{k}{k}$ of $\SO{3} \times \SO{3}$, and not in the representations $\mbf{k}{l}$ with $k \neq l$. Moreover, while the overall Kaluza-Klein spectrum shows increasing masses with level $n$, see figure \ref{f:Spectrum}, the masses of the tachyonic modes seem to remain stable, neither decreasing nor increasing, with increasing level $n$, as shown in figure \ref{f:Tachs}. This seems to generate a sort of ``mass gap'' just above the BF bound. However, this may be an artefact of only working with the first few Kaluza-Klein levels. Indeed, if we assign the states to Kaluza-Klein towers, as shown in figure \ref{f:Towers}, we find that each Kaluza-Klein tower eventually has increasing masses with increasing level $n$. Therefore, we expect that at a high enough level, there will be no more tachyonic states and the total number of tachyons in the Kaluza-Klein spectrum is finite.

Indeed, we expect the total number of tachyonic modes to be finite due to the fact that the scalar mass operator on a compact manifold is a self-adjoint elliptic operator. More specifically, it is a generalised Lichnerowicz operator which in the case at hand is obtained by computing the second metric variation of the internal part of the bosonic action of $D=11$ supergravity around the vacuum corresponding to the uplift of the $\SO{3} \times \SO{3}$ critical point. After choosing a gauge (see e.g. \cite{Biran:1983iy} for the corresponding analysis around the maximally supersymmetric SO(8) vacuum), this operator has the same principal symbol as the Laplace-Beltrami operator for the given metric, hence a discrete spectrum with no accumulation point \cite{Baer}. What is remarkable, however, is that the mass spectrometer formula of \cite{Malek:2019eaz,MalekSamtleben} allows us to explicitly follow the ``flow'' of the low-lying eigenvalues as the vacuum is deformed from the maximally supersymmetric one to the SO(3)$\,\times\,$SO(3) vacuum, and that the eigenvalues are not uniformly lifted up, but exhibit a crossover behaviour.

\paragraph{Moduli} At level 2 we find an additional 27 massless scalar fields on top of the Goldstone scalars. These infinitesimal moduli of the solution transform in the $3 \cdot \mbf{3}{3}$ representation of $\SO{3} \times \SO{3}$. These modes, therefore, correspond to infinitesimal perturbations of the AdS$_4$ vacuum which preserve the AdS$_4$ part but deform the compactification in a way which breaks the $\SO{3} \times \SO{3}$ symmetry. At levels 3 -- 6, all massless scalar fields are Goldstone modes, suggesting that these 27 modes are the only moduli.

It would be interesting to determine if these infinitesimal moduli can be integrated into finite deformations. Since this is a non-supersymmetric vacuum, already the existence of infinitesimal moduli suggests some hidden structure. This same structure may also protect the moduli from obstructions at higher orders. If these modes can be integrated to finite moduli, the non-supersymmetric AdS$_4$ vacuum studied here would belong to a family of AdS$_4$ vacua with even fewer symmetries. It is unlikely that the other members of the family, if it exists, would admit a consistent truncation to 4-dimensional supergravity so that the other members can only be studied directly in 11 dimensions.

\paragraph{} Using the fluctuation Ansatz of \cite{Malek:2019eaz,MalekSamtleben} it is possible to determine the 11-dimensional fields corresponding to the tachyons and infinitesimal moduli we identify. This would allow us to study the possible endpoint of the instability and whether there are obstructions to integrating up the moduli to finite deformations.

\section{Conclusions} \label{s:Conclusions}
In this paper, we computed the Kaluza-Klein spectrum of the non-supersymmetric $\SO{3} \times \SO{3}$-invariant AdS$_4$ vacuum of 11-dimensional supergravity \cite{Godazgar:2014eza}. We showed that some of the higher Kaluza-Klein modes (starting at level 2 above the four-dimensional supergravity level) violate the BF bound and thus trigger an instability. Therefore, even though this AdS vacuum is stable within the maximal four-dimensional $\SO{8}$ gauged supergravity, it is not stable within 11-dimensional supergravity. Our result presents the first example of tachyonic Kaluza-Klein modes and dramatically confirms the AdS swampland conjecture \cite{Ooguri:2016pdq} holds for this vacuum, which has long stood out as a possible counterexample.

In particular, we found that higher Kaluza-Klein modes can have masses below the lowest-lying modes, contradicting the observation based on supersymmetric solutions. Another interesting feature of the Kaluza-Klein spectrum is that we found 27 massless scalar fields at level 2 which are not eaten by massive vector fields and, therefore, the $\SO{3} \times \SO{3}$-invariant AdS$_4$ solution has infinitesimal moduli which break the $\SO{3} \times \SO{3}$ symmetry. Finally, this paper demonstrates the power of the methodology developed in \cite{Malek:2019eaz,MalekSamtleben} which allowed us to perform the first computation of the Kaluza-Klein spectrum of a supergravity solution whose cover is not spin.

There are several questions about the $\SO{3} \times \SO{3}$ AdS$_4$ vacuum studied here that it would be interesting to investigate. Firstly, it would be worthwhile to better analytically understand the structure of the Kaluza-Klein spectrum. For example, why do tachyons only appear in the symmetric representations $\mbf{k}{k}$ of $\SO{3} \times \SO{3}$? Another question is to identify the $S^7$ modes that correspond to the tachyons and infinitesimal moduli we uncovered. One way of doing this is by evaluating the mass matrix \eqref{eq:ScalarMassMatrix} as a function of the deformation parameter $\lambda$ which links the round $S^7$ to the $\SO{3} \times \SO{3}$-invariant deformation studied here, see \eqref{eq:ScMSO3SO3}. A second direction of inquiry is to use the fluctuation Ansatz in \cite{Malek:2019eaz,MalekSamtleben} to uplift the Kaluza-Klein modes, whose masses we computed here, to fluctuations of the 11-dimensional solution. This would shed light on the possible end-point of the instability triggered by the Kaluza-Klein modes and also on the relationship to the brane-jet instability of this AdS$_4$ vacuum \cite{Bena:2020xxb}. Similarly, it would allow us to explore whether the infinitesimal moduli we found in the Kaluza-Klein spectrum can be integrated to finite moduli of the 11-dimensional AdS$_4$ solution. This would imply that this non-supersymmetric AdS$_4$ vacuum is part of a continuous family of non-supersymmetric AdS$_4$ vacua in 11-dimensional supergravity, which break the $\SO{3} \times \SO{3}$ symmetry.

To conclude we would like to emphasise that, despite the important progress reported in this paper, the question as to the existence or non-existence of stable non-supersymmetric AdS vacua of string and M-theory remains wide open. First of all, there are non-supersymmetric vacua in the $\mathrm{ISO}(7)$ gauged theory which are stable within the four-dimensional gauged supergravity \cite{Borghese:2012qm,Guarino:2015qaa,Guarino:2019snw,Guarino:2020jwv}. Moreover, using the consistent truncation of massive IIA supergravity on $S^6$ \cite{Guarino:2015jca,Guarino:2015vca}, these vacua can be uplifted to 10-dimensional solutions of string theory. Therefore, the method of \cite{Malek:2019eaz,MalekSamtleben} can be readily applied to address the stability of these non-supersymmetric AdS$_4$ vacua, which have also recently be shown to be brane-jet stable \cite{Guarino:2020jwv}. Moreover, there is a far richer variety of non-supersymmetric stationary points for maximal (and non-maximal) gauged supergravities in three dimensions than for higher dimensions $D\geq 4$, whose stability remains to be investigated. For instance, the maximal gauged $\SO{8} \times \SO{8}$ gauged theory of \cite{Nicolai:2001sv} possesses more than 2700 critical points \cite{FPrivate}, among them at least one stable non-supersymmetric one \cite{Fischbacher:2002fx}. However, a major unsolved problem here concerns the possible uplifts of these vacua to 10 or 11 dimensions, as these cannot correspond to standard Kaluza-Klein compactifications. There are also other string models with broken supersymmetry, and which seem to show stability \cite{Basile:2018irz}.

\section*{Acknowledgements}
The authors thank Christian B\"{a}r, Nikolay Bobev, Thomas Fischbacher, Adolfo Guarino, Krzysztof Pilch, Alessandro Tomasiello, Daniel Waldram and Nick Warner for helpful discussions. EM and HN  are supported by the ERC Advanced Grant ``Exceptional Quantum Gravity'' (Grant No. 740209).

\appendix

\section{SU(8) {\em vs.} SL(8,$\mathbb{R}$) Bases}\label{app:SU8SL8}
We can express a vector $\cT_M$ in the $\mathbf{56}$ representation of $E_{7(7)}$ in terms of antisymmetric tensors in the $\SL{8,\mathbb{R}}$ and $\SU{8}$ bases as follows. Under $\SL{8,\mathbb{R}} \subset E_{7(7)}$, the $\mathbf{56}$ of $E_{7(7)}$ decomposes as $\mathbf{56} \longrightarrow \mathbf{28} \oplus \mathbf{28}'$ and accordingly we can write
\begin{equation}
 \cT_M = \left( \cT_{ab},\, \cT^{ab} \right) \,,
\end{equation}
where $a, b = 1, \ldots 8$ are used for the $\SL{8,\mathbb{R}}$ basis and $\cT_{ab} = - \cT_{ba}$ and $\cT^{ab} = - \cT^{ba}$ are real. On the other hand, under $\SU{8} \subset E_{7(7)}$, $\mathbf{56} \longrightarrow \mathbf{28} \oplus \bar{\mathbf{28}}$ and accordingly we have
\begin{equation}
\cT_M = \left( \cT_{ij},\, \cT^{ij} \right) \,,
\end{equation}
where now $i, j = 1, \ldots, 8$ are used for the $\SU{8}$ basis and $\cT_{ij} = - \cT_{ji}$ and $\cT^{ij} = - \cT^{ji}$ are complex but related by complex conjugation, i.e.
\begin{equation}
 \cT^{ij} = \bar{\cT}_{ij} \,.
\end{equation}
The relationship between these two basies is analogous to the relationship between $\SU{1,1}$ and $\SL{2,\mathbb{R}}$ \cite{Cremmer:1978ds}
\begin{equation} \label{eq:56SU8SL8}
\begin{pmatrix} {\mathcal{T}}_{ij} \\  {\mathcal{T}}^{ij} \end{pmatrix} \,\equiv\,  \frac1{4\sqrt{2}}
\Gamma^{ab}_{ij}  \begin{pmatrix} {\mathcal{T}}_{ab} + i{\mathcal{T}}^{ab}
\\ {\mathcal{T}}_{ab} - i{\mathcal{T}}^{ab} \end{pmatrix} \,,
\end{equation}
using the $\SO{8}$ $\Gamma$-matrices $\Gamma^{ab}_{ij}$ with the normalisation
\begin{equation}
 \Gamma^{ab,ij}\, \Gamma_{cd,ij} = 16\, \delta^{ab}_{cd} = 8 \left( \delta^a_c \delta^b_d - \delta^a_d \delta^b_c \right) \,.
\end{equation}
Similar relations hold for other representations of $E_{7(7)}$ by writing them as tensor products of the $\mathbf{56}$ representation and acting on each $\mathbf{56}$ as in \eqref{eq:56SU8SL8}.

For completeness, we will also give the relationship between the scalar field fluctuations in the $\SU{8}$ and $\SL{8,\mathbb{R}}$ bases. In the $\SU{8}$ basis, the scalar Kaluza-Klein modes are parameterised, as in \eqref{eq:ScalarSU8}, by complex self-dual four-forms of $\SU{8}$, i.e.
\begin{equation}
j^{ijkl}{}_{\Sigma} = \bar{j}_{ijkl,\Sigma} = \frac{1}{4!} \epsilon^{ijklmnpq} j_{mnpq,\Sigma} \,,
\end{equation}
where $\bar{j}_{ijkl,\Sigma}$ is the complex conjugate of $j_{ijkl,\Sigma}$, and $\epsilon^{ijklmnpq}$ is the eight-dimensional alternating symbol. On the other hand, in the $\SL{8}$ basis the $j_{AB,\Sigma}$ are parameterised by a real symmetric traceless tensor, $\phi_{ab,\Sigma}$, and a real self-dual four-form $\phi_{abcd,\Sigma}$ as follows
\begin{equation}
 \begin{split}
  j^{ab}{}_{cd,\Sigma} &= - j_{cd}{}^{ab}{}_{\Sigma} = \delta^{[a}{}_{[c} \phi^{b]}{}_{d],\Sigma} \,, \\
  j^{abcd}{}_{\Sigma} &= j_{abcd,\Sigma} = \phi_{abcd,\Sigma} \,,
 \end{split}
\end{equation}
where
\begin{equation}
 \phi_{ab,\Sigma} = \phi_{ba,\Sigma} \,, \qquad \phi^a{}_{a,\Sigma} = 0 \,, \qquad \phi_{abcd,\Sigma} = \frac{1}{4!} \epsilon_{abcdefgh} \phi^{efgh}{}_{\Sigma} \,,
\end{equation}
and all indices are raised/lowered by $\delta_{ab}$.

The relation between the real scalar fields $\phi_{ab,\Sigma}$ and $\phi_{abcd,\Sigma}$ in the SL(8,$\mathbb{R}$) basis and the complex self-dual field $j_{ijkl,\Sigma}$ is given by
\begin{equation}
j_{ijkl,\Sigma} = \frac1{16}\, \Gamma^{ab}_{ij} \, \Big( \phi_{ac,\Sigma}\, \delta_{bd} + i \phi_{abcd,\Sigma} \Big)\, \Gamma^{cd}_{kl} \,,
\end{equation}
so that for instance
\begin{equation}\label{JJ1}
 j^{ijkl,\Sigma}\, j_{ijkl,\Sigma} = \frac32 \phi^{ab,\Sigma}\, \phi_{ab,\Sigma} + \phi^{abcd,\Sigma}\, \phi_{abcd,\Sigma} \,.
\end{equation}

\providecommand{\href}[2]{#2}\begingroup\raggedright\endgroup

\end{document}